\documentclass[conference]{IEEEtran}
\IEEEoverridecommandlockouts
\usepackage{cite}
\usepackage{amsmath,amssymb,amsfonts,bm}
\usepackage{algorithm,float}
\usepackage{algpseudocode}
\usepackage{graphicx,subfigure}
\usepackage{textcomp}
\usepackage{xcolor}
\usepackage{multirow}
\usepackage{balance}
\usepackage{colortbl}
\usepackage{url}
\usepackage{caption}
\def\BibTeX{{\rm B\kern-.05em{\sc i\kern-.025em b}\kern-.08em
    T\kern-.1667em\lower.7ex\hbox{E}\kern-.125emX}}

\definecolor{org}{rgb}{0.9,0.4,0.1}

\begin{document}
	\bstctlcite{IEEEexample:BSTcontrol}
	\title{SVD-Embedded Deep Autoencoder for\\ MIMO Communications}
	\author{%
		\IEEEauthorblockN{Xinliang Zhang$^{\dagger}$, Mojtaba Vaezi$^{\dagger}$, and Timothy J. O’Shea$^{\ddagger}$ }
		\IEEEauthorblockA{
			$^{\dagger}$Department of Electrical and Computer Engineering,
			Villanova University, Villanova, PA, USA\\
			$^{\ddagger}$Virginia Tech and DeepSig, Inc., Arlington, VA, USA\\
						Emails:  \{xzhang4, mvaezi\}@villanova.edu$^\dagger$, oshea@vt.edu$^\ddagger$
		}
	}
	\maketitle

\begin{abstract}
	Using a deep autoencoder (DAE) for end-to-end communication in multiple-input multiple-output (MIMO) systems is a novel concept with significant potential.  
	DAE-aided MIMO has been shown to outperform  singular-value decomposition (SVD)-based 
	precoded  MIMO in terms of bit error rate (BER). 
	This paper proposes	embedding  left- and right-singular vectors
		of the channel matrix  into DAE encoder and decoder to further  improve the 
		performance of the MIMO DAE. 
		SVD-embedded DAE largely outperforms theoretic linear 
		precoding in terms of BER. 
		This is remarkable since it demonstrates that  DAEs have  
		significant 
		potential to exceed the limits of current system design  by 	
		treating the communication 
		system as a single, end-to-end optimization block. Based on the simulation 
		results, at 
		SNR=10dB, the proposed SVD-embedded design can achieve a BER of about $10^{-5}$  and  reduce the BER at 
		least 10 times compared with existing 
		 DAE without SVD, and up to 18 times  
		compared with theoretical linear precoding. We attribute this to the 
		fact that the proposed DAE can match the input and output as an 
		adaptive modulation structure with finite alphabet input. 
		We also observe that adding residual connections to the DAE further improves the  performance.
\end{abstract}

\section{Introduction}\label{sec_intro}
The {fundamental functionality}  of communication 
systems is to effectively transmit a message 
from a source to a destination over a channel. A well-established principle of 
communication systems design is to 
break up the system into multiple blocks, each with a specific 
responsibility, for instance, source coding, channel coding, precoding, 
modulation, and equalization, etc., and design them separately.  {The fact 
	that} each block therein can 
be independently designed and optimized {makes the system design 
	simpler, but} such a split-up optimization is not 
{globally} optimal at the system level \cite{o2017introduction}.

Deep learning (DL)-based {end-to-end} communications system 
\cite{o2017introduction}, on the other hand,  provides a new perspective to 
communications system design and aims to jointly optimize the 
components at the transmitter  and receiver  without any handcraft block structure.  
Unlike today's system design which carries decades of research and implementation efforts, 
DL-based communication is still in its infancy stage. 
{The simplicity and potential of this new  system 
design  approach are appealing and motivate us to exploit the potential of the
end-to-end communications system design in the physical layer.}

Recent attempts on DL in the physical layer for the single-input single-output 
channel have shown that,  as a refreshing 
communication 
system design method, \textit{autoencoder} (AE) can jointly optimize encoding, decoding, and signal 
representation \cite{o2017introduction,dorner2017deep, 
	ye2020deep}. 
{AE is an artificial neural network that learns to match its 
output to its input.  It has an encoder and decoder and 
tries to reconstruct the original input by
minimizing the reconstruction error often through some impairment such as lossy or compressive representation.}
So, AE is an unsupervised 
learning 
approach. As such, it is intriguing how it improves the overall performance of 
communication systems in terms of bit error rate (BER). 
A deep autoencoder (DAE) ia an AE with a deep neural
network.  DAEs can be used for   
transmitter and receiver design. In \cite{o2017introduction},  it is shown that 
DAE reaches the same block 
error rate as binary
phase-shift keying (BPSK) modulation with Hamming (7,4) code and 
outperforms uncoded BPSK. 
The authors in \cite{dorner2017deep} implemented the DAE on hardware. 
 DAE can also 
significantly outperform convolutional codes  in the high signal-to-noise 
ratio (SNR) \cite{ye2020deep}.

DAE design is more challenging when it comes to multiple-input 
multiple-output (MIMO) systems.  {DAE in MIMO 
communication systems was first investigated in \cite{o2017deep} where
	a MIMO spatial}
multiplexing system using DAE is designed with the perfect knowledge of 
channel state information at the transmitter (CSIT).  The DAE takes 
advantage of  CSIT to cancel  interference at the receiver, and can be 
seen as decorrelator/interference nuller  the same role singular-value 
decomposition 
(SVD) plays in  MIMO systems \cite{tse2005fundamentals}. 
{Later, the authors in \cite{song2020benchmarking} applied channel state information 
at the 
	receiver (CSIR), which  improves 
	the performance in terms of symbol error rate (SER).
	


\subsection{Contributions} 
{In this paper, we focus on DAE design for MIMO 
	communication with perfect CSIR and CSIT. We introduce an
	SVD-embedded DAE which improves the state-of-the-art DAE in terms of BER.}
	The contributions of this paper can be summarized as follows:
	\begin{itemize}
		\item We propose and design an SVD-embedded DAE network. 
		Precoding 
		and pre-processing at the transmitter and receiver have been added as one layer with 
		non-trainable weights. {This design can achieve BER nearly 
		$10^{-5}$ at SNR=10dB,} and can reduce the BER {10 to 
			30 times}  compared with  the plain DAE without any precoding layers.  
		The encoder of SVD-embedded DAE uses the transmitted bits/symbols 
		along with the CSIT as inputs, and its decoder exploits both the 
		received symbols and the CSIR. 
		\item Remarkably, {the designed DAE outperforms the {SVD 
		precoding} MIMO system from 1.5 to 18 times on BER 
			 {depending on the SNR value}.} 
		We attribute this amazing result to the fact that the SVD-embedded DAE 
		can optimize and match the input and output symbols like an adaptive 
		modulation structure {to fit the precoding and presumably the target 
		information rates corresponding to each eigenvalue/eigenmode 
		channel,} whereas the conventional methods use a 
		block-based system optimization structure, which is known to be
		sub-optimal even if we know optimal solutions in each 
		block. 
		

		\item  {We test the performance of per-bit input and per-bit output regression and compare it with the commonly-used one-hot input and one-hot output  \cite{o2017deep,song2020benchmarking}}. From our 
		simulation, bit input is a good candidate for DAE especially when the 
		number of input bits is 2 and 4.  On the other hand, the one-hot 
		input has a great potential for 6-bit input at high SNR. 
		\item We add residual connections, i.e., shortcuts, in the proposed DAE. 
		This design helps improve the performance, especially for the bit inputs.  
		Instead of widening the network, the shortcuts can increase the {learning capacity and performance of the network}  without extra parameters. It keeps the network with previous 
		layer parameters, which may reduce the effect of the vanishing gradient 
		problem. {A skip connection helps preserve gradient and allows it to learn representations at different depths. }
	\end{itemize}
}

\addtolength{\topmargin}{0.01in}
\subsection{Related Work}
{The DL 	has been exploited for   improving the performance of many physical layer 
	communication components, such as channel coding 
	\cite{farsad2018deep},  channel modeling \cite{ye2018channel},  
	precoding 	\cite{zhang2021multi, shi2021deep},  and massive MIMO \cite{shi2021deep, 
	wen2018deep}. Besides,  DAE for end-to-end communication has attracted increasing interest in 
various communications problems.  DAE, as a semi-supervised learning strategy, can handle scenarios  
with limited or 
 no training data set and still can outperform existing  solutions.  
For example,  in
\cite{felix2018ofdm},  an orthogonal frequency division 
multiplexing  based DAE system is developed which is robust against 
synchronization and multipath channel. DAE is used for channel coding 
design in \cite{jiang2019turbo}, which can achieve near-optimal 
performance.}
In general, DL is a  suitable  learning and optimization tool and  
	has a high potential for improving the computational complexity, and as a result can 
	improve energy efficiency  when such systems are deployed.
	
	The remainder of this paper is organized as follows. We elaborate on the 
	MIMO system model 
	in 
	Section~\ref{sec_sys}. The MIMO DAE design is 
	introduced in Section~\ref{sec_mimo}.
	We then present the training approach in Section~\ref{sec_train}, numerical 
	results in  Section~\ref{sec_simulation}, and 
	conclude the paper in Section~\ref{sec_con}.

	\section{System Model}\label{sec_sys}
	\subsection{MIMO System}
\begin{figure}[h]
	\centering
	\includegraphics[width=0.49\textwidth]{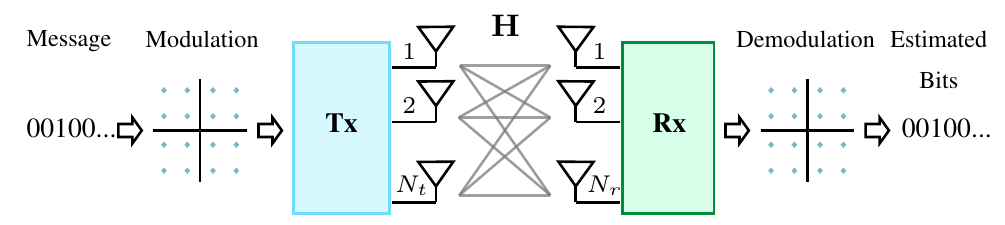}
	\caption{High-level system model of  MIMO wireless 
	communications. In this paper, we consider $N_t=N_r=2$.}
	\label{fig_sys} 
\end{figure}

	A simplified digital communications system consists of a transmitter (Tx) and 
	receiver (Rx) pair communicating over a wireless channel, as shown in 	
	Fig.~\ref{fig_sys}. 
	{The ultimate goal is to transmit information bits and receive them with a certain BER. The binary message is first converted to symbols using a modulation scheme.}
	Assume the Tx and Rx  are  equipped with $N_t$ and $N_r$ antennas, respectively.
	{The Tx wants to send to the Rx the information vector $\mathbf{s}$ which contains $N_s$ bits. To do so, the $N_s$ bits in $\mathbf{s}$ are encoded to a transmit sample (or named signal vector) $\mathbf{x} \in 
	\mathbb{C}^{N_t \times 1}$ and sent through the $N_t$ antennas. The Tx has an average power 
	constraint $E\{\mathbf{x}\mathbf{x}^{H}\} \leq P$. 
	} 
	The
	received signal at the Rx can be  expressed as
	\begin{align}\label{eq_recSig}
		\mathbf{y} = \mathbf{H}\mathbf{x} + \mathbf{w},
	\end{align}
	{where  ${\mathbf{H}} \in 
	\mathbb{C}^{N_r \times N_t}$  is the Rayleigh flat-fading channel  between the Tx and Rx  whose elements follow 
	$\mathcal{CN}(0,1)$}
	and $\mathbf{w} \in \mathbb{C}^{N_r\times 1}$ 
	is  additive  Gaussian white noise (AWGN) {with  mean zero and variance of $\sigma_N^2$}. The Rx 
	is supposed to  
	decode the received signal $\mathbf{y} $ and estimate the transmitted bit vector $\hat{\mathbf{s}}$. {In this paper, 
	we focus on the case that both the Tx and Rx have two antennas.}
	

\begin{figure*}[t]
	\centering
	\includegraphics[width=0.99\textwidth]{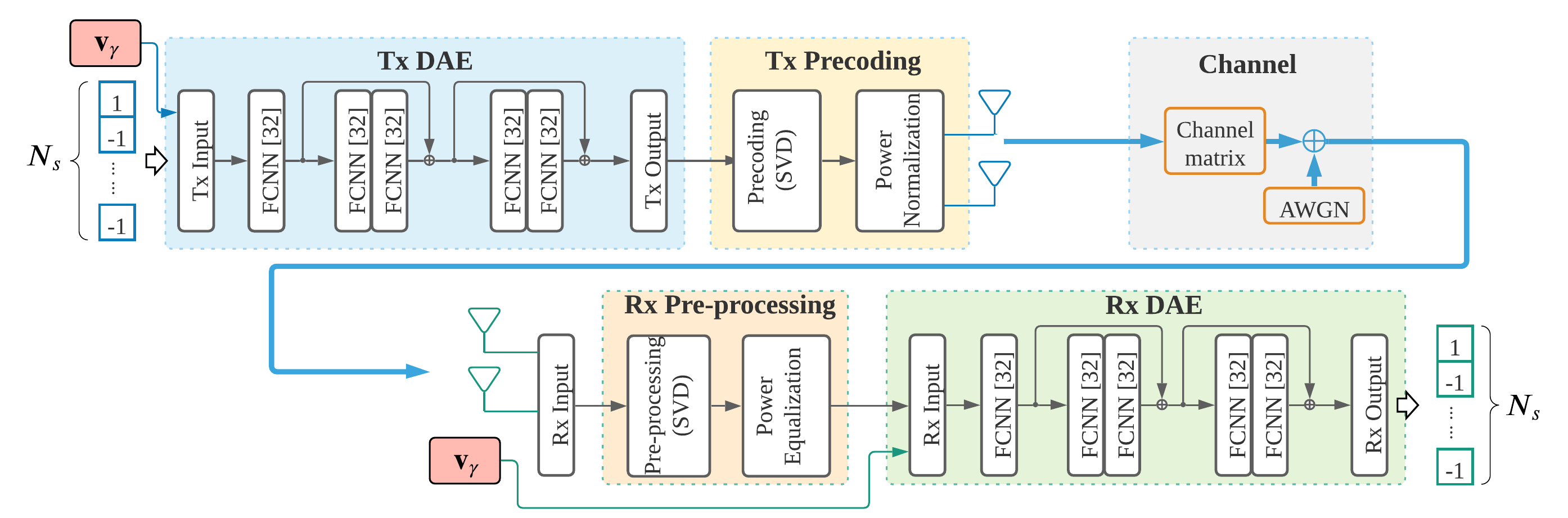}
	\caption{The proposed complex DAE architecture with SVD. 
		$\mathbf{v}_\gamma$ contains the singular values of the channel 
		matrices and it is given  to both of the Tx DAE and Rx DAE.}
	\label{fig_NET1} 
\end{figure*}
\subsection{SVD and Water-Filling Algorithm}\label{ssec_svd}
To exploit spatial multiplexing in MIMO systems,  precoding and power 
allocation have been developed to eliminate the {inter-channel interference} at each 
antenna at the Rx and to increase the data rate. SVD-based precoding 
\cite{cover2012elements} {which is the optimal  precoding in a closed-loop MIMO system}, designed by utilizing CSIT and CSIR. In this 
method, the channel ${\mathbf{H}}$ is decomposed as 
\begin{align}\label{eq_svd}
	\mathbf{H}=\mathbf{U}\mathbf{\Lambda}\mathbf{V}^H,
\end{align}
{where $\mathbf{U}$ and 
$\mathbf{V}$ are two unitary matrices and  $\mathbf{\Lambda}$ is the 
diagonal matrix with the singular values of ${\mathbf{H}}$. The matrices 
$\mathbf{V}$ and $\mathbf{U}^H$ are applied as precoding and 
pre-processing matrices. Then we have the observation as
\begin{align}
\tilde{\mathbf{y}}\triangleq\mathbf{U}^H\mathbf{y}
=\mathbf{\Lambda\Sigma}\mathbf{s}+\mathbf{U}^H\mathbf{w},
\end{align}
where $\mathbf{s}$ is the symbol vector that each element is independent 
and identically distributed (i.i.d.) symbol, 
$\mathbf{x}\triangleq\mathbf{V}\mathbf{\Sigma}\mathbf{s}$ is the signal and 
also the result of power allocation and precoding, $\mathbf{\Sigma}$ is the 
power allocation matrix obtained by water-filling (WF) \cite{cover2012elements}. In this way, SVD precoding  
eliminates the 
interference at each antenna and converts the correlated MIMO into a set of 
parallel sub-channels whose gain is the singular value of the channel matrix and the
square root of the power allocated. }

\addtolength{\topmargin}{0.01in}
\section{SVD-Embedded DAE} \label{sec_mimo}
The proposed SVD-Embedded DAE, (named as SVD-DAE for simplifying),  
structure for point-to-point (P2P) MIMO is shown in Fig.~\ref{fig_NET1}. 
In 	general, 
the system contains five parts: bit input, Tx, channel, Rx, 
and bit output. The Tx includes  AE and precoding which 
construct a transmitted signal. Rx includes pre-decoding and  
AE to decode the received signal. 

\subsection{Input Design}\label{ssec_input}
{The input of the SVD-DAE contains two parts. The first part is for 
the information to be transmitted--the $N_s$ bits. The second 
part, $\mathbf{v}_\gamma$, is a feature vector related to the channel state information (CSI). The 
details of the two parts are described below.}
\subsubsection{Transmitted Information}
We feed the network with a vector containing $N_s$ bits in each 
transmission. 	Each bit is represented by $\pm 1$. Two input structures 
are considered: 
\begin{itemize}
	\item 
	{\textit{Bit input}}:
The bit vector $\mathbf{v}_b$ is sent to the \textit{Tx DAE}  directly. The 
number of 
elements in $\mathbf{v}_b$ is $N_s$, which means the number of bits in each 
transmission period. In each period we transmit one  $\mathbf{v}_b$, all the 
bits are independent and identically distributed following Binomial 
distribution. 
\item 
{\textit{One-hot input}}: The one-hot input has already been used in  
\cite{o2017deep, song2020benchmarking}. However, 
typically, the information to be transmitted is in bits. In this case, the bit 
vector is initially converted to a one-hot vector $\mathbf{v}_o$ by 
Gray coding and then transmitted to the DAE. Essentially, the encoding could 
be done arbitrarily since the DAE does not affect the order of the one-hot 
vectors.
\end{itemize}
\subsubsection{CSI Feature}
Besides the bits for transmission, another feature vector $\mathbf{v}_\gamma$ is added which contains the effect of 
channel matrix and noise power level. 
{We define $\bm\gamma$ as the channel gain to noise ratio, where the $i$th element in $\bm\gamma$ 
for $i$th sub-channel as ${\gamma}_i\triangleq\frac{{\lambda}^2_i}{\sigma_N^2}$. 
Then, we define the feature vector  
as} 
\begin{align}\label{eq_snrFeature}
	\mathbf{v}_\gamma\triangleq S_c
	\left(
	0.2\times 10
	\log_{10}({\bm\gamma}+10^{-6})
		\right),
\end{align}	
where $S_c(t)$ is the zero-centered Sigmoid function 
rectifying the ${\rm\gamma}$ in dB between $-1$ and $1$, the factor $0.2$ 
scales  ${\rm\gamma}$ in dB, and $10^{-6}$ is used to prevent 
${\rm\gamma}$ going to zero and $\log_{10}({\bm\gamma})$  to
 negative infinity.

To summarize, the bit input vector is composed as 
$\mathbf{v}_I=[\mathbf{v}_b, \mathbf{v}_\gamma]$ containing 
$N_s+N_t$ values, whereas for one-hot input, $\mathbf{v}_I=[\mathbf{v}_o, 
\mathbf{v}_\gamma]$, the size is $2^{N_s}+N_t$.

\subsection{\textit{Tx DAE} Design}\label{ssec_txAE}
The encoder  of  the SVD-DAE consists of  a fully-connected 
neural network (FCNN) with two shortcut connections 
\cite{he2015deepResidual}. The  size of the input is equal to the 
length of $\mathbf{v}_I$ while the output size is $N_t$. Each of the five layers 
in the middle  contains 32 hidden nodes. The shortcuts, also called the 
residual connections, are used to provide a direct connection from the output 
to the input, which enables to convey the gradient in back-propagation and 
reduce the difficulty in 
training \cite{ji2019convae}.

\subsection{\textit{Tx Precoding} Design}\label{ssec_txPrecoding}
In the \textit{Tx Precoding} part, we add one layer with non-trainable weights 
that will not be updated during the training process. The 
weights are set according to right-singular vectors of the channel matrix, 
which is the optimal precoding for P2P MIMO 
problem with an infinite alphabet. When the channel coefficients are 
obtained, the weights of the \textit{Tx Precoding} layer are set as 
$\mathbf{W}_{\rm T}=\mathbf{V}$ given in \eqref{eq_svd}. The 
bias in each layer
is set to be zeros. 
Before the transmission, a normalization layer 
linearly adjusts the total power at the output of \textit{Tx Precoding}. The 
power of the transmit signal is equal to $P$. {In this paper, $P=20$W is fixed for all DAEs and every signal  transmission (per-example). 
}
\subsection{Channel and Noise}\label{ssec_channel}
The channel is formed by a non-trainable layer between the Tx and Rx. The 
weight of the channel layer is the channel matrix, i.e., $\mathbf{W}_{\rm 
	Ch}=\mathbf{H}$. White Gaussian noise is added after implementing the 
channel layer. For each sub-channel, the noise power is $\sigma_N^2$. Since  
the bandwidth of  the streams on each antenna {is equal to the 
 sampling frequency, the noise power then equals the noise 
spectral density, $\sigma_N^2=N_0$.} To measure the overall signal and 
noise level, we define SNR following \cite{song2020benchmarking} that
\begin{align}\label{eq_EB}
{\rm SNR} =\frac{P}{N_0}= N_s\frac{E_b}{N_0}, 
\end{align}
where $E_b\triangleq \frac{P}{N_s}$ refers to the energy per bit. The influence 
of the channel is not accounted for in this definition.

\subsection{\textit{Rx Pre-processing} Design}\label{ssec_rxPrecoding}
In this part, we add two layers with non-trainable weights and zero biases. 
The number of hidden nodes is $N_t$. The weights in the first layer are 
 $\mathbf{W}_{\rm R1}=\mathbf{U}^H$, where $\mathbf{U}^H$ is given in 
 \eqref{eq_svd} and is left-singular vectors of the channel coefficients. 
{Since $\mathbf{U}^H$ is a unitary 
	matrix  and we assume the noise on each receiving antenna has the same 
	power,   $\mathbf{U}^H\mathbf{w}$ has the same power as $\mathbf{w}$. 
	So 
	the noise level  is  not changed.} Together with  
the \textit{Tx Precoding}, an equivalent parallel channel can be obtained 
from the input of \textit{Tx DAE} to the output of \textit{Rx 
	DAE}. The second layer in \textit{Rx Pre-processing} unifies the 
diagonals to 1, whose weights are the pseudo-inverse of  
$\mathbf{\Lambda}$, i.e., $\mathbf{W}_{\rm R2}={\rm 
	pinv}(\mathbf{\Lambda})$.

\subsection{\textit{Rx DAE} Design}\label{ssec_output}
The input of \textit{Rx DAE} contains the output of \textit{Rx 
	Pre-processing} and SNR feature, i.e., $\mathbf{v}_\gamma$ given in 
\eqref{eq_snrFeature}. The network structure, number of hidden layers, and 
nodes in each layer are the same as those in \textit{Tx DAE}. The 
number of  \textit{Rx DAE} output varies according to the input 
design in Section~\ref{ssec_input}. When using bit inputs, the output size is 
$N_s$. Finally, the outputs 
are  converted to bit vector $\hat{\mathbf{v}}_b$ and the BER for one 
transmitted vector is 
\begin{align}\label{eq_ber}
\epsilon_b= \frac{1}{N_s}\sum_{\hat{\mathbf{v}}_b(j)\neq\mathbf{v}_b(j)}1. 
\end{align}
{For one-hot inputs, the output size is $2^{N_s}$. The output 
$\hat{\mathbf{v}}_o$s are firstly demodulated using Gray code. Then we 
calculate BER in a similar way as \eqref{eq_ber}.} 

{The loss function for {bit input} and {one-hot input} are  
mean-square error (MSE) and cross-entropy respectively. In general,  
cross-entropy is more efficient for training since a high loss value can be 
obtained on errors. However, it can only work for one-hot, a classification-like 
output. On the contrary, MSE is suitable for both regression
and classification, while the cross-entropy is limited.}

\begin{table}[t]
	\centering
	\caption{Parameters of the proposed DAE}
	\label{tab_DAE}
	\begin{tabular}{c|c|l}
		\hline
		\multirow{2}{*}{Name}     & Hidden 
		& \multicolumn{1}{l}{If the weights are trainable} \\
		&layers$\times$nodes&\multicolumn{1}{l}{and the weights (if not)}
		\\ \hline
		Tx Precoding      & $1\times N_t$ &  N,  
		$\mathbf{W}_{\rm T}=\mathbf{V}$, $\mathbf{b}_{\rm T}=\mathbf{0}$  
		\\ \hline
		Channel             & $1\times N_t$ &  N,  
		$\mathbf{W}_{\rm Ch}=\mathbf{H}$, $\mathbf{b}_{\rm Ch}=\mathbf{0}$
		\\ \hline
		\multirow{2}{*}{Rx Pre-processing} 
		& $1\times N_t$ & N,  
		$\mathbf{W}_{\rm R1}=\mathbf{U}^H$, $\mathbf{b}_{\rm 
			R1}=\mathbf{0}$
		\\ 
		&   $1\times N_t$ & N,      
		$\mathbf{W}_{\rm R2}={\rm pinv}(\mathbf{\Lambda})$, 
		$\mathbf{b}_{\rm 
			R2}=\mathbf{0}$   \\ \hline
	\end{tabular}
\end{table}

The parameters of the proposed DAE are summarized in 
Table~\ref{tab_DAE}.
The main difference of the proposed DAE  with the existing 
architecture in {\cite{song2020benchmarking}} is the two 
{new blocks: \textit{Tx Precoding} appended after \textit{Tx DAE} and \textit{Rx 
	Pre-processing} before \textit{Rx DAE}}. The precoding and 
pre-processing introduced in this design is the optimal solution, SVD,  for 
P2P MIMO precoding with the infinite alphabet {\cite{telatar1999capacity}}.  
We name the proposed architecture 
SVD-embedded DAE and use the shortened name \textit{SVD-DAE}. If the SVD precoder and pre-processing are excludes, i.e., if we let $\mathbf{W}_{\rm 
T}=\mathbf{W}_{\rm 
	R1}=\mathbf{W}_{\rm R2}=\mathbf{I}$, then we call it \textit{plain-DAE}.   


\section{Training Process}\label{sec_train}
In the training session, we first randomly generate  $M_c$ channel 
matrices whose elements follow i.i.d. complex Gaussian distribution 
$\mathcal{CN}(0,1)$. In this paper, we consider $P=20$W and $N_s\in\{1,2,4,6\}$ bits. For each $N_s$, we have a different DAE instance which is trained  with a stochastic strategy: 
\begin{itemize}
\item We first train the DAE with the first channel and the batch size is 
	$M_b$.
{\item In each batch, $N_0$ is uniformly and randomly chosen from $-20$dB to $25$dB. Both channel and $N_0$ are fixed. 
\item After applying the weights using Adam optimizer,
we train the DAE with the next channel and another $N_0$.}
\item We continue until all $M_c$ channels are involved in the training, and the first round of training is finished. 
\item Next, the learning rate $l_r$ is reduced by a factor $\alpha_r$  and 
	DAE goes to the next round of training. 
\end{itemize}
{The weights of the DAE are updated once per channel, 
and the loss of the DAE is expected to reduce to fit the channel. By repeatedly 
traversing all the channels, the DAE is trained in a stochastic manner. 
Theoretically, the DAE will adapt to an arbitrary channel if $M_c$ goes to 
infinity and traverses all possible channel matrices.
 In this paper, we set $M_c=2000$ and  the number of rounds is 
$M_r=1000$. 

The training approach is summarized in Algorithm~\ref{alg_Train}. 
{The detailed 
	$E_b/N_0$ 
	regions in the training stage are given in Table~\ref{tab_DAE2}. 
In the test stage, the DAEs are  evaluated over $[-10, 
20]$dB. }

\begin{algorithm}[h]
	\caption{Training Procedure}\label{alg_Train}
	\begin{algorithmic}[1]
		{
			\State Select $N_s$, the number of bits in each transmission. 
			\State Set $M_c=2000$, $M_b=2000$, and $M_r=1000$;
			\State Set $l_r=10^{-4}$, $\alpha_r=0.995$, {and $P=20$W};
			\State Initialize the DAE network.
			\For {index $m_r$ from $1$ to $M_r$}
			\For {index $m_c$ from $1$ to $M_c$}
			\State Randomly and uniformly set $N_0$ in $[-20, 25]$dB;
			\State Calculate and set the weights and biases in DAE
			\State $\quad$according to Table~\ref{tab_DAE};
			\State {Randomly  generate $M_b$ bit vectors. Each bit 
			fol-}
			\State {$\quad$lows Binomial distribution with 
			probability}
			\State {$\quad$$P_b=0.5$}; 
			\State Train the DAE once with batch-size $M_b$; 
			\EndFor
			\State Update: $l_r=\alpha_r l_r$;
			\EndFor
		}  
	\end{algorithmic}
\end{algorithm}


\begin{table}[h]
	\centering
	\caption{SNR and $E_b/N_0$ region during the 
		training according to \eqref{eq_EB}. $N_0$ is uniformly distributed.}
	\label{tab_DAE2}
	\begin{tabular}{c|cccc}
		\hline
		{$N_s$} & {$P$ (W)} & {$N_0$ (dB)}    & {SNR (dB)}     & {$E_b/N_0$ (dB)} 
		\\ \hline
		{1}     & 20      & [-20, 25] & [-12,33] & [-12, 33]  \\
		{2}     & 20      & [-20, 25] & [-12,33] & [-15, 30]  \\
		{4}     & 20      & [-20, 25] & [-12,33] & [-18, 27]  \\
		{6}     & 20      & [-20, 25] & [-12,33] & [-20, 25]  \\ \hline
	\end{tabular}
\end{table}



\begin{figure*}[t] 
	\centering
	\subfigure[$N_s$=1]{
		\includegraphics[scale=.57]{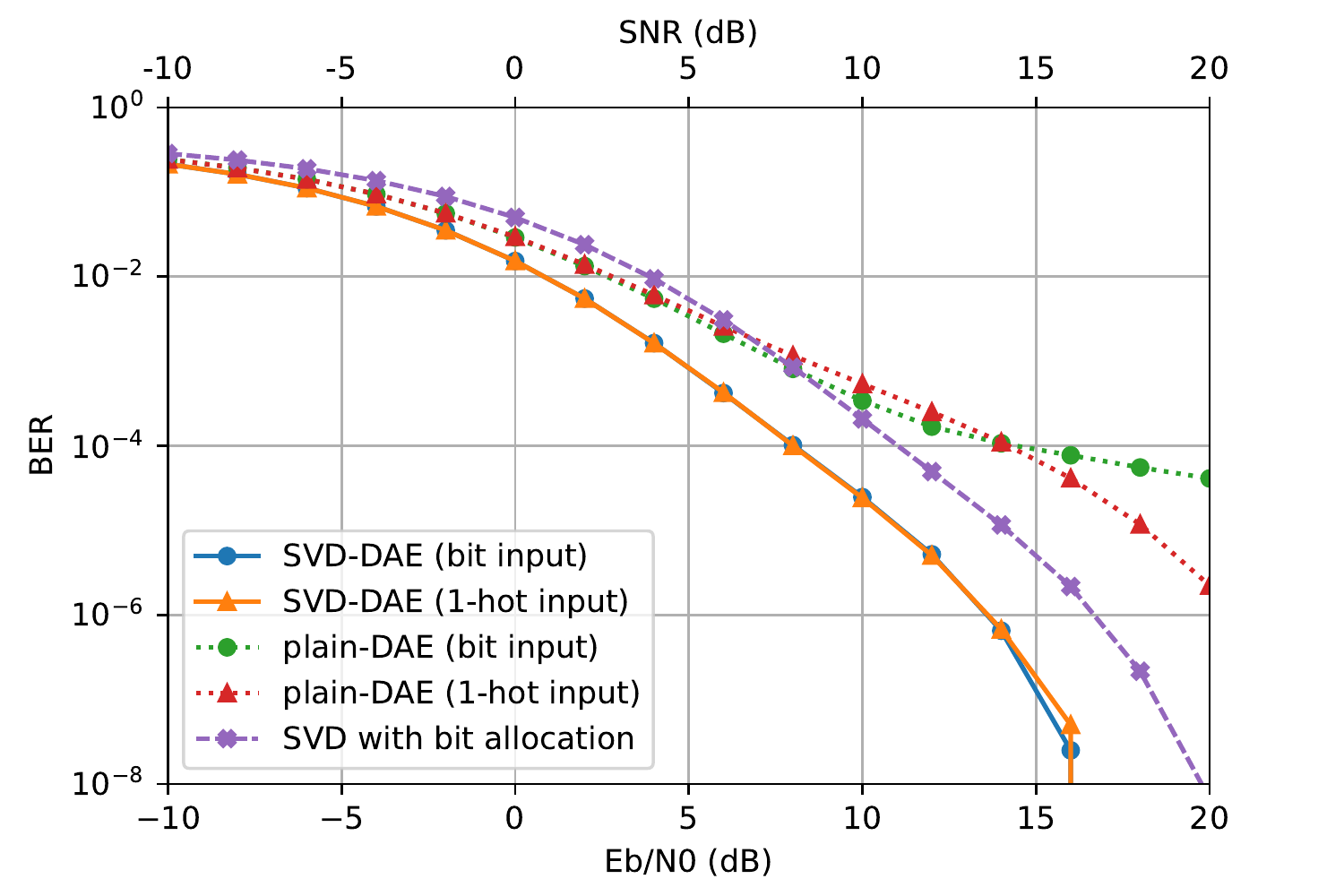}
		\label{fig_comp_svd1a} 
	}
	\subfigure[$N_s$=2]{
		\includegraphics[scale=.57]{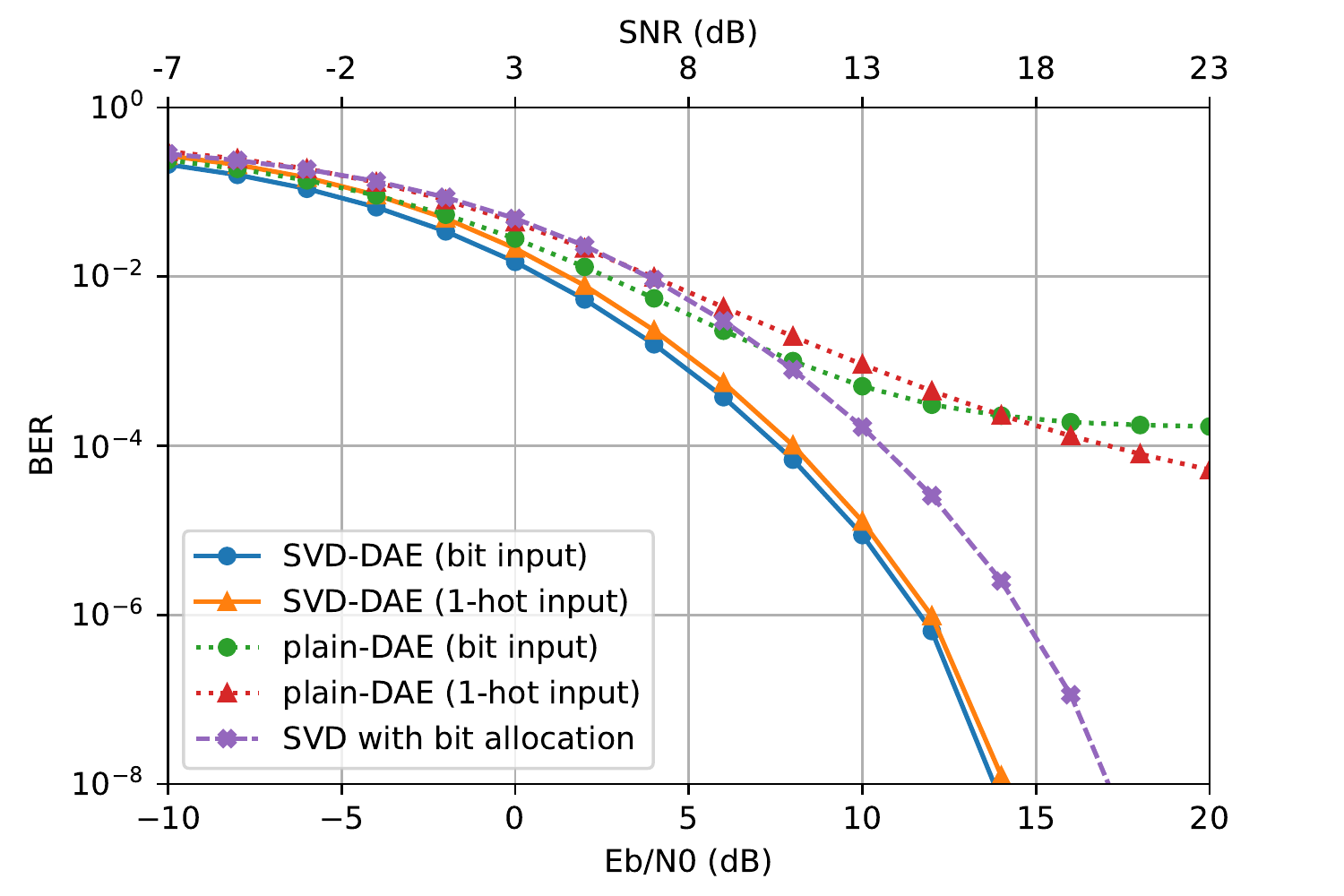}
		\label{fig_comp_svd1b} 
	}\\
	\subfigure[$N_s$=4]{
		\includegraphics[scale=.57]{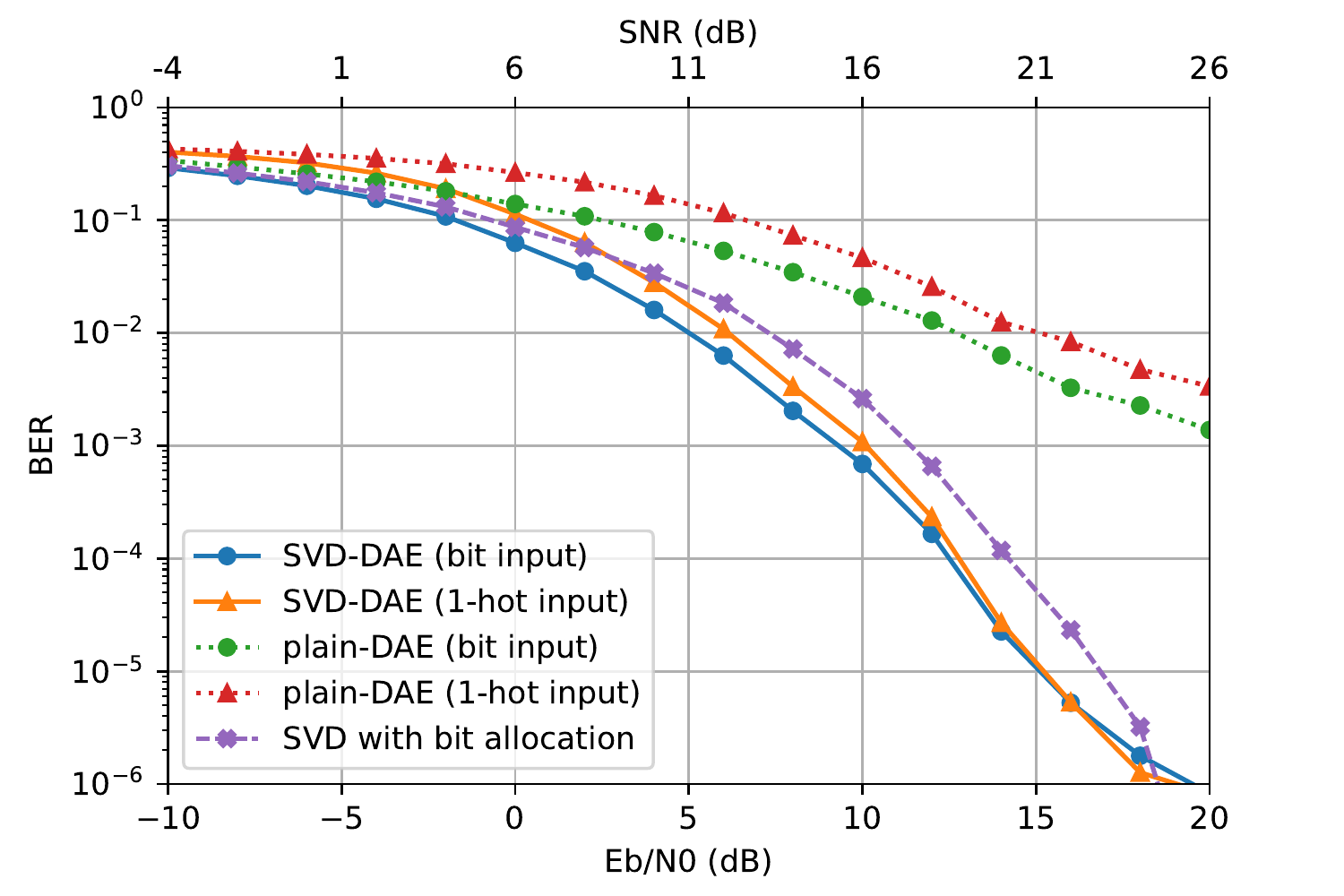}
		\label{fig_comp_svd1c} 
	}
	\subfigure[$N_s$=6]{
		\includegraphics[scale=.57]{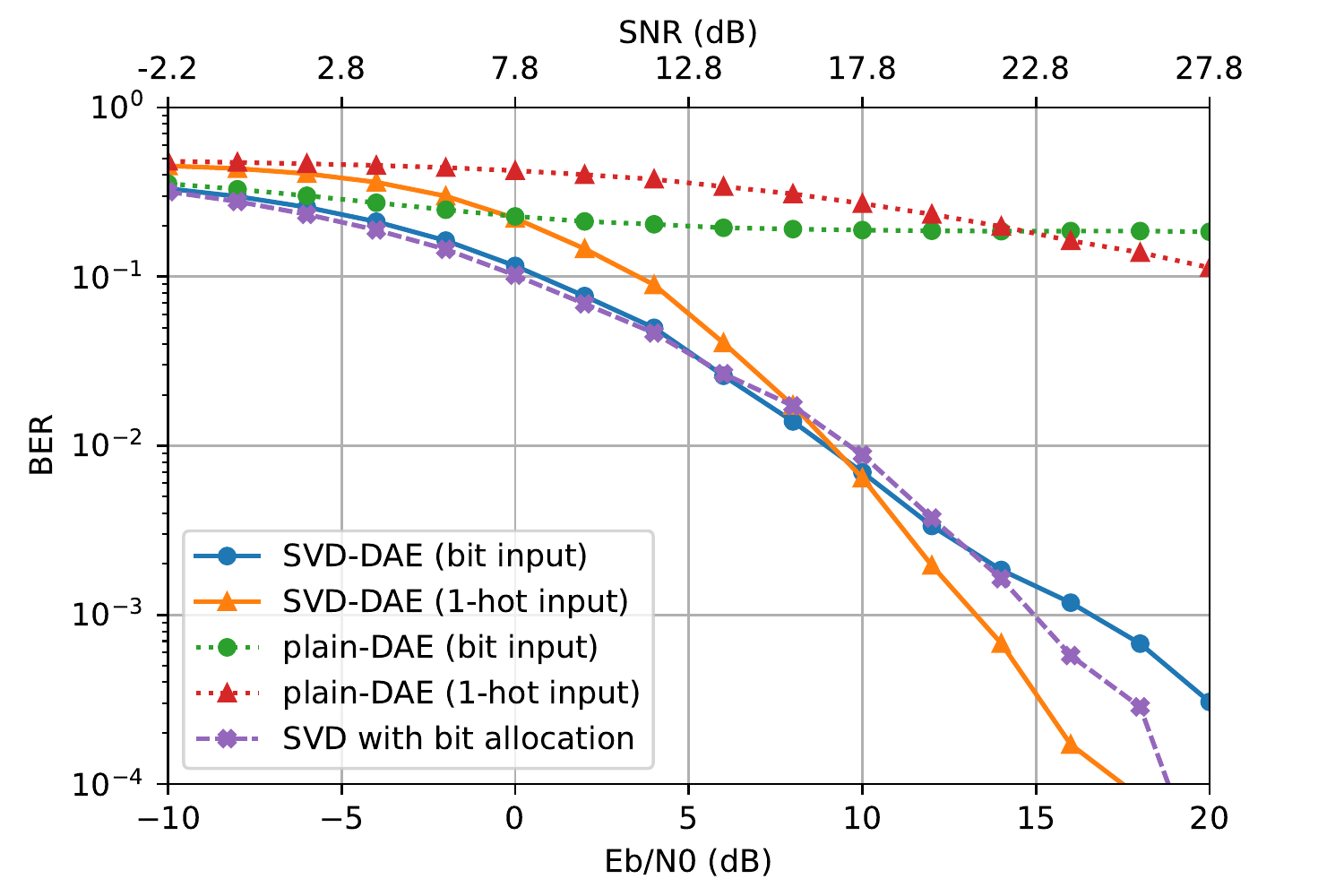}
		\label{fig_comp_svd1d} 
	}
	\caption{BER comparison between SVD-DAE and plain-DAE. 
		The top x-axis denotes the SNR (dB) and the bottom x-axis denotes the 
		$E_b/N_0$ (dB), where the relation is shown in \eqref{eq_EB}. All the 
		DAEs 
		have residual connections as in Fig.~\ref{fig_NET1}.}
	\label{fig_3levels} 
\end{figure*}
\section{Numerical Results}\label{sec_simulation}

\subsection{Performance of SVD in DAE}

We first examine the performance of having the two new blocks, Tx Precoding 
and Rx Pre-processing, in the DAE. {We plot the BER versus  
$E_b/N_0$ and SNR in each figure. The horizontal axis  on the bottom of 
each figure and  SNR is on the top. 
The relation between SNR and $E_b/N_0$ is given in \eqref{eq_EB}.} For each 
$E_b/N_0$, all the neural networks are tested and average over 2000 
randomly generated channels. 

The results in Fig.~\ref{fig_3levels} show the comparison between SVD-DAE  
and plain-DAE. {The plain-DAE has the same design as in Fig.~\ref{fig_NET1} 
	excluding 
	the precoding and pre-processing parts. Besides,   the
	extra feature vector for the plain-DAE becomes a vector with the real 
	and imaginary parts of the channel matrix.} We also plot the curves for the SVD with bit 
allocation as a baseline.  Similar to \cite{tse2005fundamentals, 
o2017deep} we all have parallel sub-channels. The difference is we 
choose the best modulation schemes among BPSK and $M$-quadrature 
amplitude modulation ($M$-QAM) to reach the minimum BER over all of the 
sub-channels while the total bits transmitted is $N_s$. {For the baselines, 
we limit the average power to $P=20$W. The peak power can be larger than 
$P$ when using $M$-QAM with $M>4$.}

In general, SVD-DAE outperforms plain-DAE, that is, the solid lines 
are lower than  the dotted lines in Fig~\ref{fig_3levels}. With $N_s$ 
increasing from Fig.~\ref{fig_comp_svd1a} to  Fig.~\ref{fig_comp_svd1d}, the 
benefit of SVD-DAE compared with plain-DAE is largely enhanced. 
For 
example, {the BER reduces 30 times} when $N_s=6$ and 
$E_b/N_0=10$dB but  {10 times}  when 
$N_s=1$ and $E_b/N_0=10$dB. On the other hand, when transmitting fewer 
bits, the 
BER is smaller. {For instance, when SNR is equal to 10dB, the BER 
of SVD-DAE in Fig.~\ref{fig_comp_svd1a} is  $10^{-5}$}, while in 
Fig.~\ref{fig_comp_svd1c} and Fig.~\ref{fig_comp_svd1d}, the BER is   
{$7\times10^{-4}$ and $6\times 10^{-3}$}, respectively. 
SVD-DAE outperforms the baseline from 1.5 to 18 times on BER 
	improvement.

The bit input is better than one-hot input  for low SNRs. 
The  curves with circle markers in Fig.~\ref{fig_comp_svd1b}, 
Fig.~\ref{fig_comp_svd1c}, and Fig.~\ref{fig_comp_svd1d} are lower than the 
 curves with triangles. This is because, when using one-hot input, 
DAE cannot optimize the bit-to-symbol conversion which is outside the DAE 
and is fixed by using Gray code. {Putting it differently, the bit labeling (bit-to-symbol mapping) is 
invisible using the one-hot method. However, if we use bit inputs, the bit labeling can be 
optimized in the training.} Besides, in Fig.~\ref{fig_comp_svd1d}, when SNR is 
greater than 10dB, the SVD-DAE with one-hot input performs better than the 
bit 
input. {That is, the advantage of cross-entropy loss appears in 
one-hot input. Also, if we aim at optimizing the SER, one-hot input will 
always be better than the bit input (for saving space, we omit that result).}
In Fig.~\ref{fig_comp_svd1a}, SNR and ${E_b}/{N_0}$ are the same and 
the 1-bit input and one-hot input  achieve similar performance. 
{This 
is because  one-hot input has no BER loss in the bit-to-symbol conversion 
when 
the input  carries only one bit of information ($N_s=1$).}

\begin{figure}[t]
	\centering
	\includegraphics[width=0.49\textwidth]{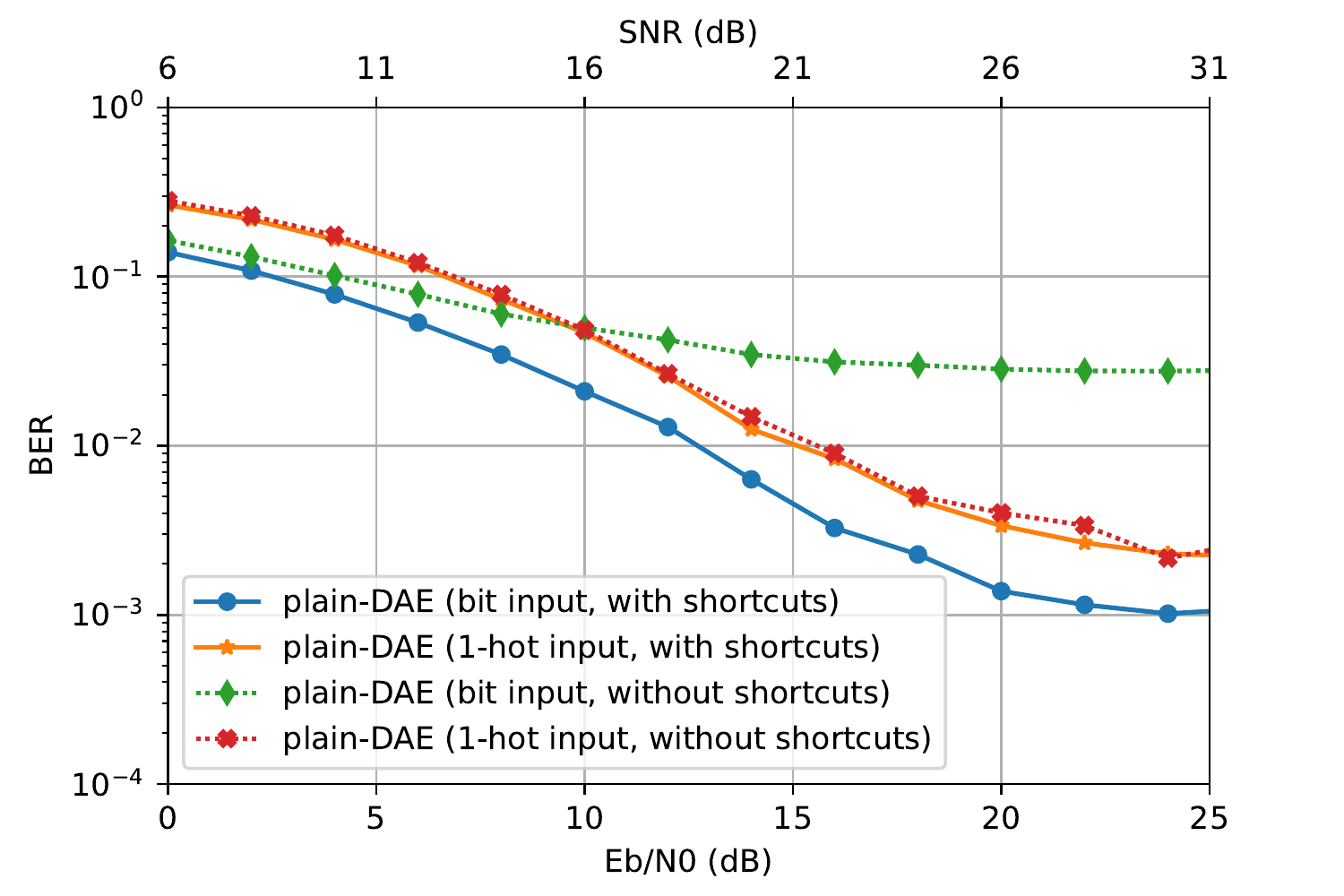}
	\caption{Performance of residual connections on plain-DAE when 
		$N_s$=4.}
	\label{fig_comp_res2a} 
\end{figure}
\begin{figure}[t]
	\centering
	\includegraphics[width=0.49\textwidth]{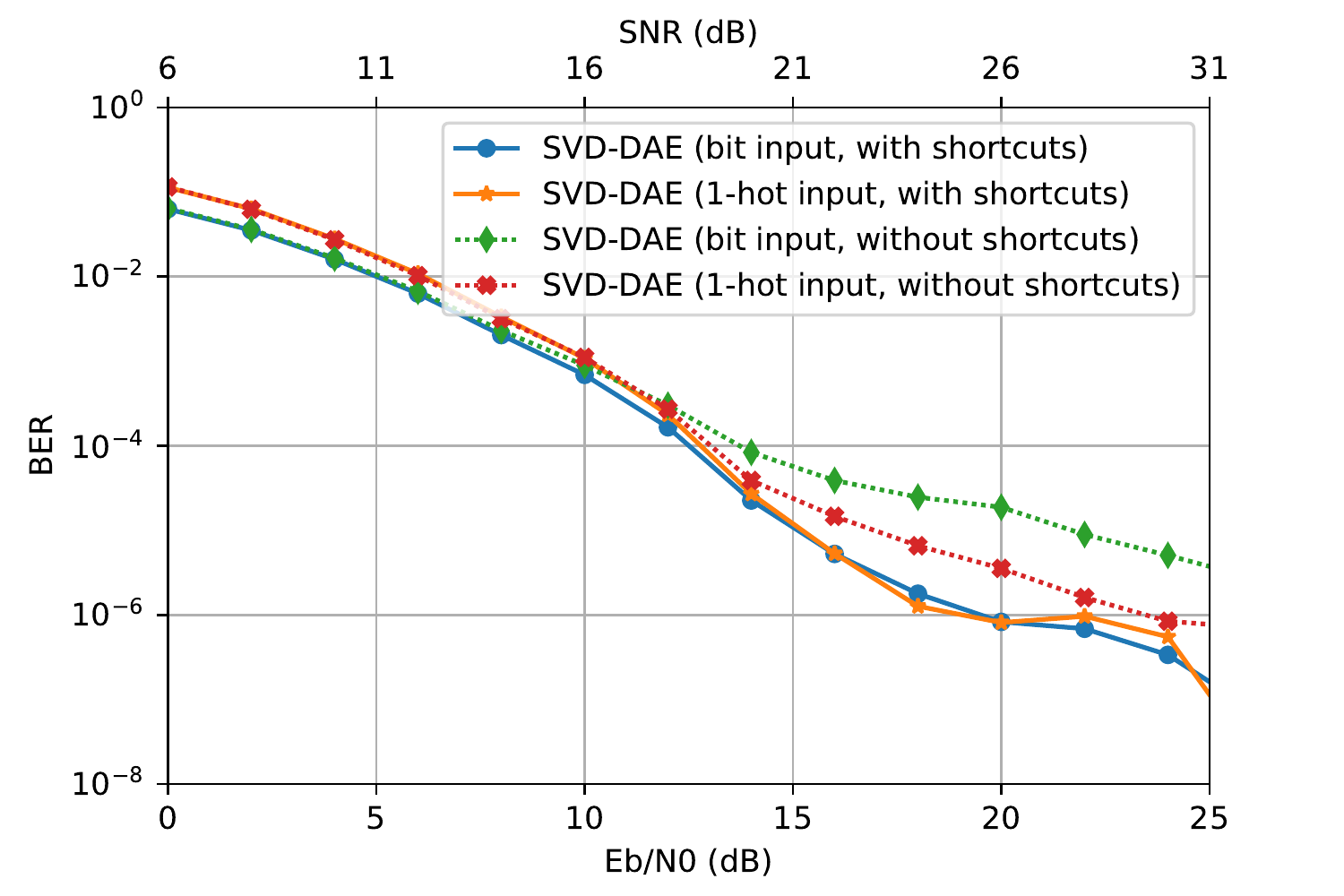}
	\caption{Performance of residual connections on SVD-DAE when  
		$N_s$=4.}
	\label{fig_comp_res2b} 
\end{figure}

\begin{figure}[t]
	\centering
	\includegraphics[width=0.49\textwidth]{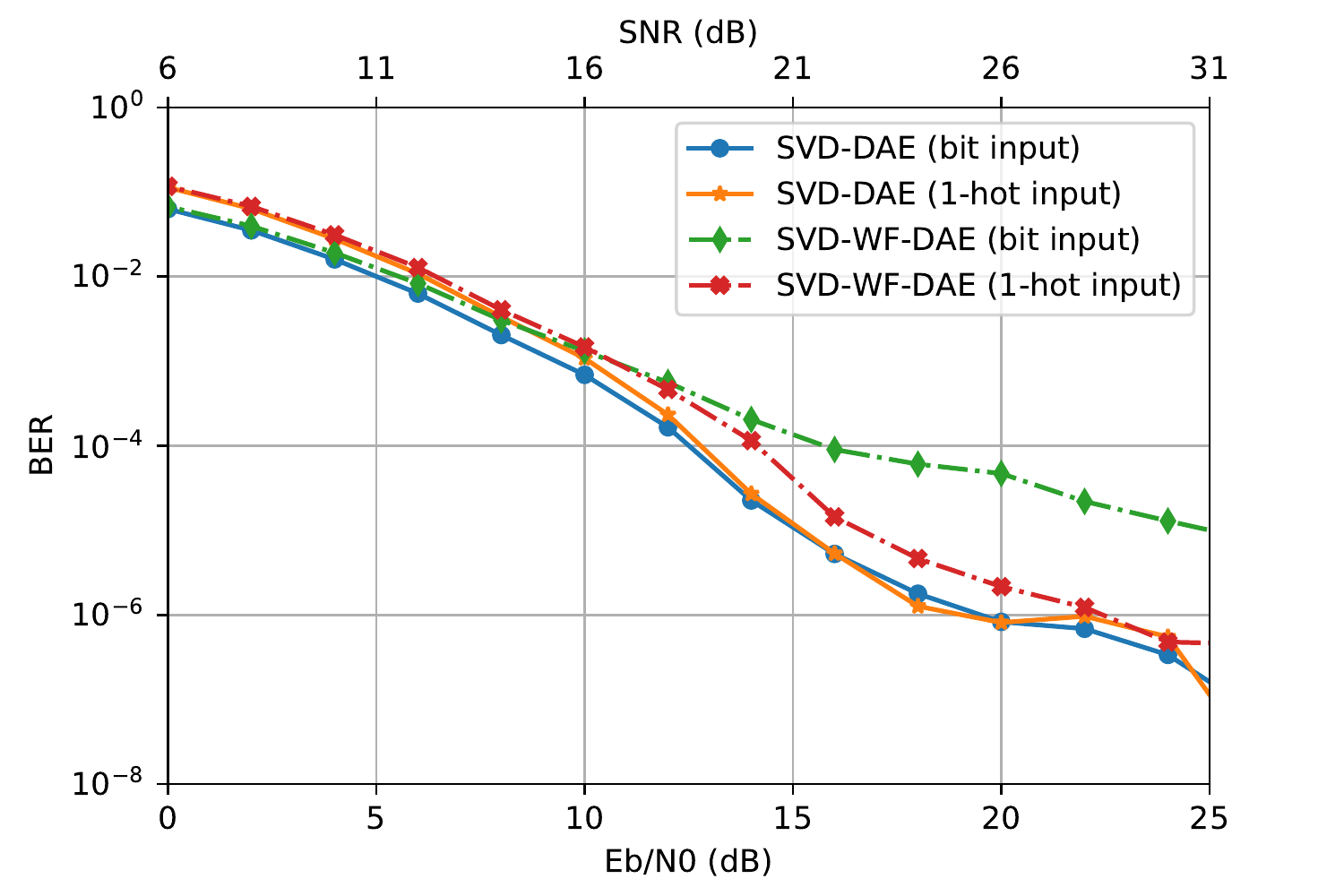}
	\caption{SVD-DAE compared with SVD-WF-DAE on BER  when $N_b$=4.}
	\label{fig_comp_WF} 
\end{figure}

\subsection{The Effect of Residual Connections in DAE}
We study the effect of adding a residual connection, i.e., shortcuts, 
{which 
enables to convey the gradient in back-propagation and reduce the difficulty 
in training by providing a direct connection from the output to the input.}
We apply the shortcuts to both plain-DAE and SVD-DAE, $N_s$ is chosen to 
be 4 for this experiment. In Fig.~\ref{fig_comp_res2a} 
and Fig.~\ref{fig_comp_res2b}, the BER is greatly reduced in both networks 
with the shortcuts compared with having no shortcuts, especially when 
$E_b/N_0$ is greater than 10dB. It reduces the BER of plain-DAE with 
bit input from {$3\times10^{-2}$  to  $10^{-3}$} when 
$E_b/N_0=20$dB (Fig.~\ref{fig_comp_res2a}). For SVD-DAE with bit inputs, 
the BER is improved from
{$2\times10^{-5}$  to  $0.9\times10^{-6}$} (Fig.~\ref{fig_comp_res2b}).   


\subsection{The Effect of Adding WF}
The SVD with water-filling algorithm (WF) is known to be optimal precoding in the MIMO system to achieve spatial 
multiplexing \cite{cover2012elements}. This encouraged us to add WF as a 
non-trainable layer to the SVD-DAE. 
{The power allocation factors given by WF can amplify or 
attenuate the output of the \textit{Tx-DAE} depending on the value greater or 
less than $1$. The 
power normalization in the 
\textit{Tx Precoding} block is also effective.
However, WF is not 
the optimal power allocation scheme in finite-alphabet MIMO systems \cite{xiao2011globally}.} This could be the reason why SVD-DAE 
without WF performs better than SVD-DAE with WF (see in 
Fig.~\ref{fig_comp_WF}). 
One explanation 
can be that the WF algorithm allocates all the power to
the stronger channel at low SNR ($E_b/N_0\leq 10$dB) which results in a good BER performance. However, when SNR increases, WF starts assigning more power to strong channels and less power to weaker channels, in which the poor channel may not make use of the power and then suffers a bad BER.  In SVD-DAE, the singular values and noise level are fed 
to the Tx and Rx DAEs. Such a design is inspired by WF and allows the AE 
networks to learn their own WF-like solution to distribute information across 
channels.



\section{Conclusions}\label{sec_con}
We have improved the existing DAE for 
MIMO systems by introducing SVD as a differentiable layer in the network. This helps DAE make better use of the inputs of Tx autoencoder to eliminate the interference and improve the performance of MIMO spatial multiplexing. 
SVD-DAE can outperform  plain-DAE, and is competitive with linear precoding solutions  in terms of BER.  
We have also investigated the ability of the proposed DAE with bit inputs and 
one-hot inputs. Based on our results, bit input is a good candidate for DAE 
especially {in reducing the BER}. By tuning the network, residual 
connections are added to increase the depth of the network and further 
improve the results. Besides, it is meaningful to investigate different 
constraints such as  per-channel, per-antenna,  per-sum-of-antennas 
average power constraints in the future.

\balance
\typeout{}
\bibliographystyle{IEEEtran}
\bibliography{REF_commu_v1.0_nolink}


\end{document}